\documentclass[fleqn,10pt]{wlscirep}
\usepackage[utf8]{inputenc}
\usepackage[T1]{fontenc}
\usepackage{amsmath}
\title{Yoga biomechanics through energy optimization: diverse kinematics and low-dimensional kinetics}

\author[1,2,*]{Sriram Sekaripuram Muralidhar}
\author[1,3]{Manoj Srinivasan}
\affil[1]{Mechanical and Aerospace Engineering, The Ohio State University, Columbus OH 43201}
\affil[2]{Biokinesiology and Physical Therapy, University of Southern California, Los Angeles CA 90007}
\affil[3]{Program in Biophysics, The Ohio State University, Columbus OH 43201}

\affil[*]{srirammu@usc.edu}

\begin{abstract}
Yoga is widely practiced for improving physical and mental health, yet the biomechanical demands of individual yoga poses remain poorly characterized. Existing experimental studies typically examine only a small number of poses or muscles, limiting their ability to guide pose selection for rehabilitation or exercise prescription. Here we present a computational framework for predicting whole-body joint torques and contact forces from two-dimensional images of yoga poses. The framework combines manual pose digitization, a sagittal-plane MuJoCo human model, and constrained metabolic energy minimization subject to static equilibrium. We applied the framework to 55 sagittally symmetric yoga poses from Light on Yoga, generating a dataset of joint kinematics, contact locations, and predicted joint torques and contact forces. These poses collectively span nearly the full physiological range of motion of the major joints while exhibiting a wide range of predicted joint torques, with the greatest average demands on the lower extremities and lumbar region. Principal component analysis revealed a striking contrast between kinematics and kinetics: nearly all ten dimensions were required to represent the joint-angle data, whereas the first three principal components captured over 90\% of the variance in joint torques. Thus, yoga poses exhibit substantial diversity in body configurations but comparatively low-dimensional patterns of predicted joint loading. Predicted joint torque rankings were also largely insensitive to uncertainty in the relative weighting of joint metabolic costs. This framework enables systematic biomechanical analysis of yoga poses and provides a foundation for quantitatively evaluating their mechanical demands and informing future rehabilitation strategies.
\end{abstract}
\begin{document}

\flushbottom
\maketitle
% * <john.hammersley@gmail.com> 2015-02-09T12:07:31.197Z:
%
%  Click the title above to edit the author information and abstract
%

\section*{Introduction}
Yoga, an ancient Indian exercise tradition, has gained global popularity since the mid-20th century \cite{singleton_yoga_2010}. The classical form of yoga, termed Hatha yoga, involves holding static (nearly isometric) poses against gravity before deliberately transitioning to the the next pose; newer forms of yoga incorporate continuous, flowing transitions between poses. Like other forms of exercise \cite{fiuza2018exercise,valenzuela2023exercise,smith2021role}, yoga has  benefits for cardiometabolic, mental, and physical health \cite{bussing_effects_2012,chu_effectiveness_2016}. Yoga interventions can improve physical well-being, relieve pain, and reduce pain medication use in people with musculoskeletal disorders \cite{mccaffrey_benefits_2012}. In people with movement disorders, including Parkinson's disease, stroke, and cerebral palsy, yoga can improve joint mobility, balance, flexibility, and lower limb strength \cite{lazaridou_yoga_2013,roland_applications_2014,thayabaranathan_understanding_2018,legault_yoga_2021}. These findings indicate yoga's therapeutic potential for physical and mental disorders. However, yoga-based therapies are not routinely prescribed in clinical practice because we lack a quantitative understanding of the biomechanical demands of individual poses, essential for making patient-specific recommendations on the type, intensity, and frequency of practice.

Previous studies of the effects of yoga poses on muscles and joints are not comprehensive \cite{westwell_evaluation_2006,chun_biomechanical_2010,yu_physical_2012,salem_physical_2013,wang_biomechanical_2013,ni_core_2014,brenneman_yoga_2015,longpre_identifying_2015,chen_biomechanical_2017,rathore_anatomical_2017,chopp-hurley_scapular_2018,kumar_musculoskeletal_2018,mullerpatan_kinematics_2019,liu_training_2021,whissell_biomechanical_2021,baum_varying_2022,deshi_electromyographical_2023,dewan_electromyographic_2023,menegaldo_kinematics_2023}. First, collectively they examine only a small fraction of the roughly 200 known poses \cite{iyengar_light_1979}. Second, individual studies focus on specific body regions (lower extremity, core, or shoulder) or selected muscle groups, partly because experimental techniques such as surface EMG are limited to superficial muscles and relatively few measurement sites.

\begin{figure}[!htb]
    \centering
    \includegraphics[width=1\textwidth]{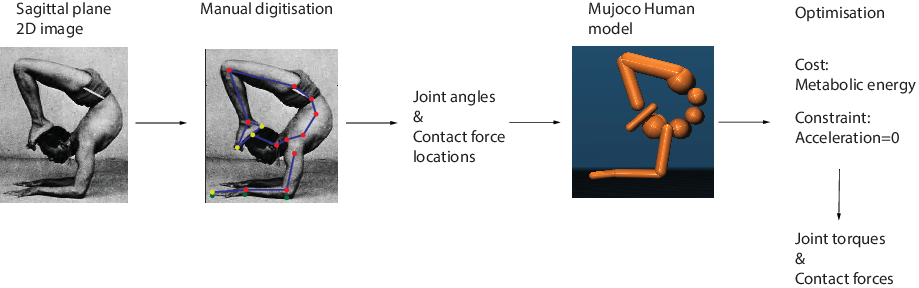}
  \caption[Simulating yoga biomechanics from yoga pose images]{\textbf{Simulating yoga biomechanics from yoga pose images.} Images of yoga poses symmetric about the sagittal plane are digitized to extract joint angles and contact locations. These are used to specify the configuration and contacts of a 2D human model in the MuJoCo physics engine. Constrained optimization minimizes metabolic energy subject to static equilibrium, predicting joint torques, joint reaction forces, and ground contact forces.}    \label{fig:frameworkExpository}
\end{figure}

Computer simulations offer a promising alternative for studying many yoga poses and their whole-body effects. Omkar et al. \cite{omkar_mathematical_2011} proposed a computational model to predict joint torques for nine two-point contact poses, but assumed negligible horizontal ground contact forces and did not validate their predictions experimentally. Here, we propose a computational framework to predict joint torques during yoga poses (figure \ref{fig:frameworkExpository}), based on the hypothesis that humans minimize metabolic energy while holding a pose \cite{srinivasan2011fifteen,brown2021unified,seethapathi2024exploration,mcallister2025behavioural,miller2014comparison,falisse2019rapid,handford2016robotic}. We demonstrate this framework by predicting joint torques for 55 classical yoga poses that are symmetric about the sagittal plane, so that their kinematics could be obtained from 2d images  \cite{iyengar_light_1979}.

\section*{Methods}
The predictive simulation workflow consists of five parts (figure \ref{fig:frameworkExpository}): (1) collecting sagittal-plane images of the poses, (2) manually digitizing the images and extract body positions and contacts, (3) extracting joint angles and contact locations from the digitized points, (4) constructing a sagittal-plane human model in MuJoCo, and (5) performing constrained optimization to minimize metabolic energy subject to static equilibrium constraints, thereby predicting joint torques, joint reactions, and contact forces.

\begin{figure}[!htb]
    \centering
    \includegraphics[width=1\textwidth]{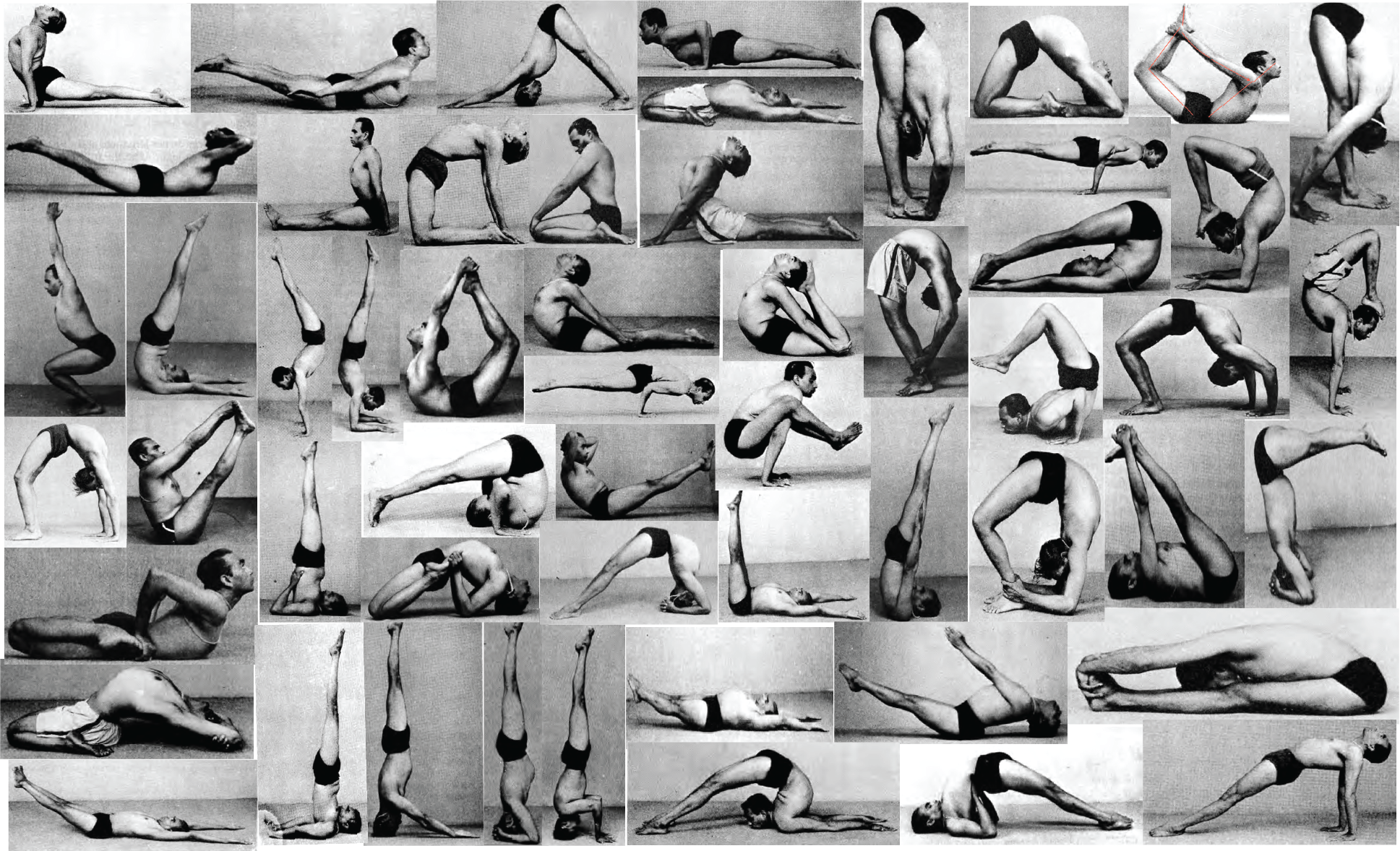}
\caption[Images of 55 yoga poses]{Sagittal-plane images of 55 yoga poses used for analysis of pose kinematics and biomechanics. Reproduced from the book \textit{Light on Yoga} \cite{iyengar_light_1979}; original photographs remain copyrighted by their respective rights holders.}
    \label{fig:allImageCollage}
\end{figure}

We analyzed 55 sagittal-plane poses from the book \textit{Light on Yoga} \cite{iyengar_light_1979} (figure \ref{fig:allImageCollage}), a modern foundational classic of yoga practice. For simplicity, we considered only poses symmetric about the sagittal plane so that they could be modeled in 2D using one arm and one leg. We therefore used \textit{Light on Yoga} \cite{iyengar_light_1979}, which contains roughly 200 poses with images and descriptions. From each pose image, we recorded the number of body segments in contact with the ground, the identities of the contacting segment pairs, and the locations of joints, the toe, heel, hand tip, and crown of the head. This contact information, absent from previous datasets \cite{suryawanshi_yoga_2023}, is essential for modeling each pose. We represented the spine, including the neck, with multiple joints (lumbar lower, lumbar mid, lumbar upper, neck) to account for spinal curvature, and modeled three joints each in the lower extremity (ankle, knee, and hip) and the upper extremity (wrist, elbow, and shoulder).

\begin{figure}[!htb]
    \centering
    \includegraphics[width=\textwidth]{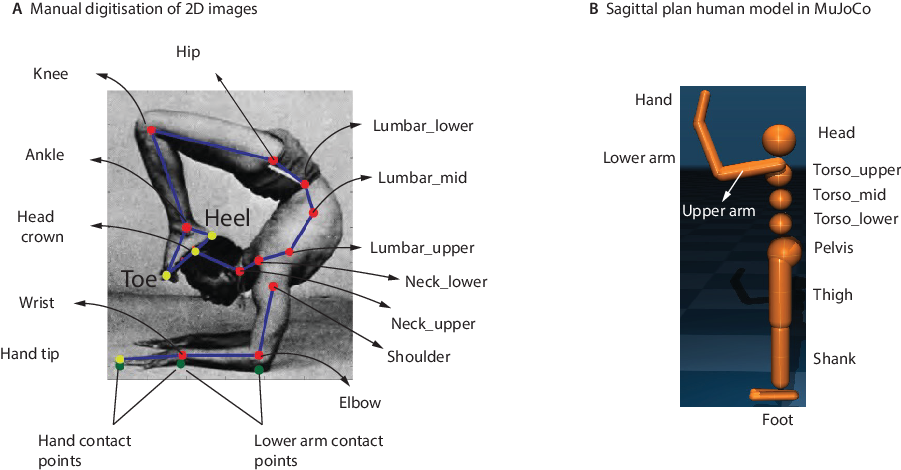}
\caption[Schematic of manual digitization and human model]{\textbf{Manual digitization and MuJoCo human model.} \textbf{A)} Digitized yoga pose showing joint centers, body extremities (toe, heel, hand tip, and crown of the head), and contact points (two each on the hand and lower arm). \textbf{B)} Sagittal-plane human model in MuJoCo with 11 rigid segments and 10 joints.}
    \label{fig:digitiseAndModel}
\end{figure}

MuJoCo (Multi-Joint dynamics with Contact) is an open-source physics engine for simulating articulated mechanical systems, including rigid-body dynamics, joint constraints, and differentiable contact interactions \cite{todorov_mujoco_2012,todorov2014convex}, and has recently found application in fast biomechanical simulations \cite{wang2022myosim}.  We performed static simulations in MuJoCo \cite{todorov_mujoco_2012}, using a 2D human model with 10 joints and 11 rigid segments, adapted from the humanoid model \cite{tassa_synthesis_2012} provided with MuJoCo (figure \ref{fig:digitiseAndModel}B). We removed joint stiffness, added a hand segment and a torso joint, and constrained all segments and joints to move in the 2D sagittal plane. The model had three joints in the upper extremity (wrist, elbow, and shoulder), three in the lower extremity (ankle, knee, and hip), and four in the spine (lumbar lower, lumbar mid, lumbar upper, and neck). We used a custom contact formulation instead of MuJoCo's native formulation, treating each segment-ground contact as a two-point contact. Contact locations were obtained from the digitized data, with one horizontal and two vertical contact force components defined for each contacting segment.

We hypothesize that humans minimize metabolic energy while holding a pose. We formulated this as a constrained optimization problem, minimizing metabolic energy subject to static equilibrium at the desired pose. In earlier work \cite{muralidhar_metabolic_2023}, we found that metabolic energy scales nonlinearly with joint torque $\tau$, specifically scaling like  $|\tau|^{\gamma}$, with $\gamma \approx 1.64$. We therefore modeled the total metabolic energy as $\sum_{i=1}^{\textrm{all joints}}\alpha_i |\tau_i|^{\gamma}$, where $\mathrm{\alpha_i}$ is a joint-specific proportionality constant. We solved for the joint torques $\tau_i$ and the contact forces $\mathbf{f}$ that minimize this metabolic objective, subject to the following constraints: the body pose $\mathrm{q} = \mathbf{q}_{\mathrm{image}}$, specified by the joint angles estimated from image digitization; zero velocity $\dot{\mathbf{q}}=\mathbf{0}$; and zero acceleration $\ddot{\mathbf{q}}=\mathbf{0}$. The accelerations were computed by MuJoCo, which implicitly solves the multibody equations of motion, $\mathbf{M}(\mathbf{q})\ddot{\mathbf{q}}+\mathbf{h}(\mathbf{q},\dot{\mathbf{q}}) =\mathbf{B}\boldsymbol{\tau} +\mathbf{J}_c(\mathbf{q})^T\mathbf{f}$.  We normal contact forces to be non-negative, allowing tangential forces without an explicit friction coefficient bound. The contacts remained stationary, consistent with static equilibrium. As a simple baseline, we assumed equal coefficients for all joints, $\mathrm{\alpha_i=1}$ but repeated the optimization 50 times with randomly generated coefficients ($\mathrm{\alpha}_i$) to assess their effect on the predictions. We solved the optimization using a trust-region algorithm \cite{nocedal2006numerical} in the SciPy Python library. The optimization predicted joint torques, joint reactions, and horizontal and vertical contact forces.

% worth putting in equations for cost and constraints

\begin{figure}[!htb]
    \centering
    \includegraphics[width=\textwidth]{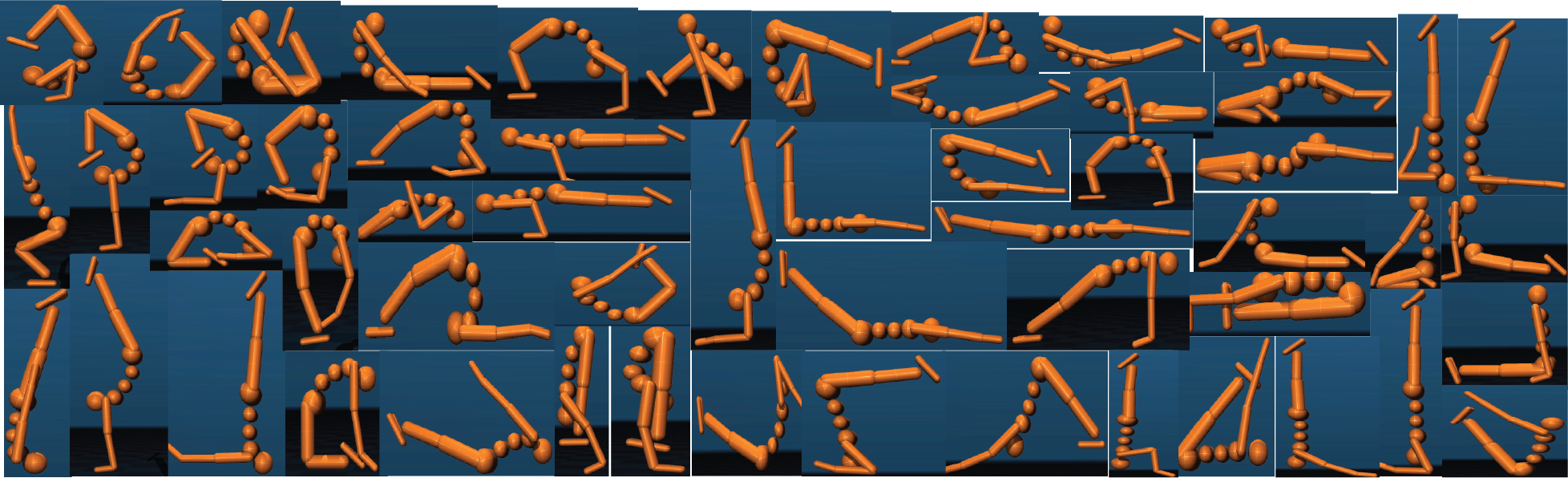}
\caption[Simulations of 55 yoga poses]{\textbf{Static simulations of 55 yoga poses in MuJoCo.} The simulated body configurations correspond to joint angles extracted from the original pose images. For each pose, constrained optimization determines the joint torques and contact forces that minimize metabolic energy while maintaining static equilibrium. All 55 optimizations converged successfully.}
    \label{fig:simulationImages}
\end{figure}

\section*{Results}
The optimizations converged for all poses, and the resulting simulations are shown in figure \ref{fig:simulationImages}. The acceleration constraint violation and optimality violation were below $10^{-4}$ for every pose (in their respective units). Each optimization required only a few seconds because the problem is static and does not require dynamic trajectory integration.

\begin{figure}[htb]
    \centering
    \includegraphics[width=\textwidth]{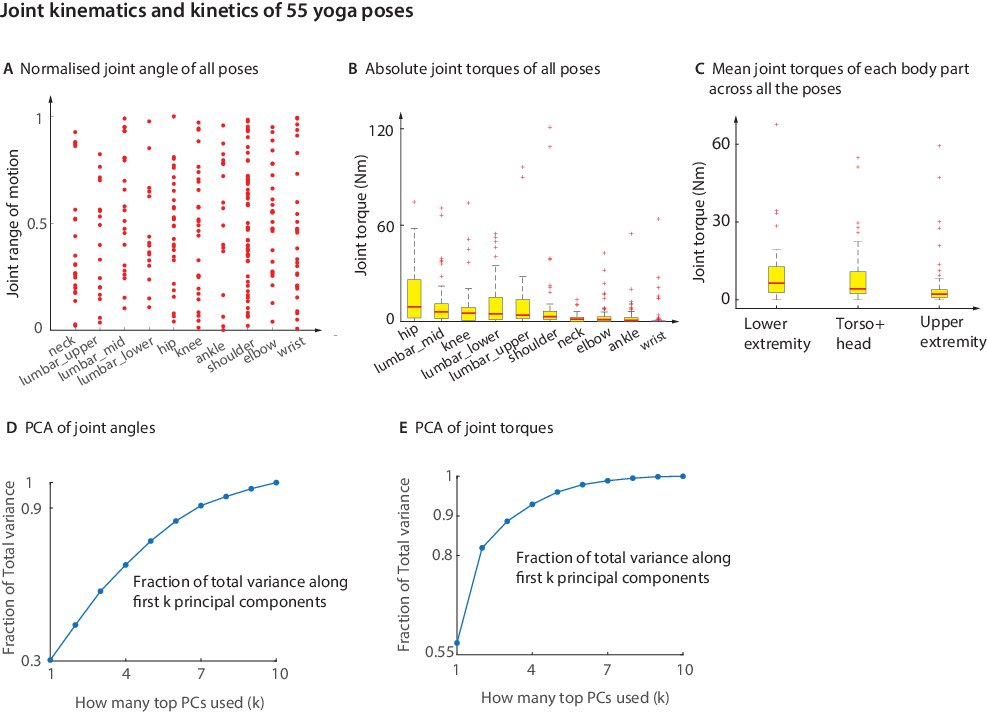}
\caption[Joint kinematics and kinetics of 55 yoga poses]{\textbf{Joint kinematics and kinetics of 55 yoga poses.} \textbf{A)} Joint angles across all poses, normalized by each joint's physiological range of motion \cite{boone_normal_1979}. Zero and one denote the minimum and maximum physiological joint angles, respectively. \textbf{B)} Absolute joint torques across all poses, predicted by metabolic energy minimization. \textbf{C)} Mean absolute joint torques for the lower extremities, torso and head, and upper extremities. \textbf{D)} Cumulative fraction of joint-angle variance explained by the first $k$ principal components. \textbf{E)} Cumulative fraction of joint-torque variance explained by the first $k$ principal components.}
    \label{fig:allPoseResult}
\end{figure}

Joint angles across the 55 poses spanned nearly the full physiological range of motion for all joints (figure \ref{fig:allPoseResult}A). We performed principal component analysis (PCA) on  the joint angles to assess the dimensionality of the poses in kinematic  space. Nearly all ten dimensions were required to represent the 55 poses (figure \ref{fig:allPoseResult}D): the first principal component explained 30\% of the variance, whereas each subsequent component explained an additional 5--10\%. Thus, the poses exhibit substantial kinematic diversity with little redundancy.

Model-predicted joint torques showed that the hip required the greatest torques on average, followed by the lumbar spine and knee, whereas the wrist required the least (figure \ref{fig:allPoseResult}B). Grouping joints by body region showed that the lower extremities experienced the greatest average torques, followed by the torso and head, and then the upper extremities (figure \ref{fig:allPoseResult}C). We similarly performed principal component analysis (PCA) on  the joint torques to assess the dimensionality of the pose repertoire in kinetic  space. In contrast to the kinematics, PCA of the joint torques showed that the first three principal components of joint torque captured over 90\% of the variance (figure \ref{fig:allPoseResult}E). Thus, despite their diverse body configurations, yoga poses exhibit comparatively low-dimensional patterns of joint loading. 

% in addition, need to also present a version of this scaled by maximum voluntary joint torque for each joint. maybe most joints will be more similar when we do that.

 % joint angles from standing position sounds strange.

\begin{figure}[htb]
    \centering
    \includegraphics[width=\textwidth]{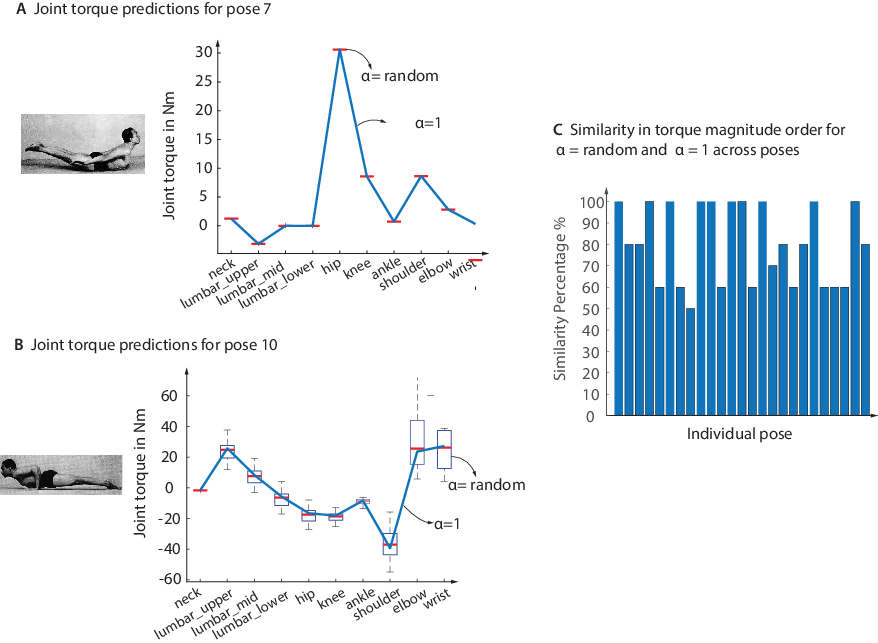}
\caption[Predictions with random coefficients $\alpha_i$ and equal coefficients $\alpha_i = 1$]{\textbf{Sensitivity of joint torque predictions to metabolic cost coefficients.} Comparison of predictions using randomly generated coefficients $\alpha_i \in (0,1)$ and equal coefficients $\alpha_i = 1$. \textbf{A)} Joint torque predictions for a representative pose insensitive to changes in $\alpha_i$.  \textbf{B)} Joint torque predictions for a representative pose sensitive to changes in $\alpha_i$.  \textbf{C)} Percentage similarity in the relative ranking of joint torque magnitudes between the median predictions across randomly generated coefficients and the predictions with equal coefficients.}
    \label{fig:alphasensitivity}
\end{figure}

Predictions were insensitive to randomly generated coefficient values ($\alpha_i$) for 29 of the 55 poses (figure \ref{fig:alphasensitivity}A). Twelve of these poses involved only a single body segment in contact with the ground, making the optimization problem fully determined. The remaining 17 poses involved multiple contacts but shared a continuous line of contact with the ground, also resulting in coefficient-insensitive predictions. The remaining 26 poses exhibited sensitivity to the coefficients $\alpha_i$ (figure \ref{fig:alphasensitivity}B). Although the predicted joint torque magnitudes varied, their relative rankings remained largely unchanged and similar to those obtained with equal coefficients ($\alpha_i=1$). The median predictions across randomly generated coefficients closely matched those from equal coefficients (figure \ref{fig:alphasensitivity}C), with 10 poses showing identical joint torque rankings and all remaining poses retaining at least 50\% similarity (figure \ref{fig:alphasensitivity}C). These results suggest that joint torque rankings are relatively robust to the choice of coefficients, even when torque magnitudes vary.

\section*{Discussion}
We developed a computational framework for predicting whole-body joint torques from 2D images of yoga poses in the sagittal plane. The framework uses a metabolic cost model to predict joint torques and contact forces under static equilibrium. It also generates a novel dataset of joint positions, contact locations, joint torques, and contact forces across 55 yoga poses. These data show that the poses collectively span nearly the full range of motion of the major joints while imposing a wide range of predicted joint torques, with the greatest average demands on the lower extremities. Principal component analysis revealed a striking contrast between kinematics and kinetics: joint angles exhibited high dimensionality, whereas the first three principal components explained over 90\% of the variance in joint torques. Thus, yoga poses are highly diverse in body configuration while producing comparatively low-dimensional patterns of predicted joint loading.

Our results are currently restricted to poses symmetric about the sagittal plane, though the framework itself does not share this restriction. Joint angles and contact locations were obtained by manual digitization and are therefore subject to measurement error. The human model consists of rigid segments and does not account for passive muscle or ligament forces or body--body contacts, which may affect predicted joint torques and forces. We also did not experimentally validate the predicted joint torques or contact forces. Finally, because the model is based on joint torques rather than individual muscles, it cannot predict individual muscle forces. Although the joint-specific metabolic coefficients are not precisely known and affected torque predictions for some poses, the relative magnitudes of joint torques were largely insensitive to their values.

Future work could extend the framework to three-dimensional poses by combining computer vision methods for estimating 3D joint kinematics with a 3D MuJoCo model. Incorporating passive joint mechanics and body--body contact would further improve biomechanical realism. Model predictions could be validated experimentally using measurements of ground reaction forces and joint torques estimated through inverse dynamics. More broadly, the dataset of joint kinematics and kinetics generated here provides a foundation for quantitatively evaluating the biomechanical demands of individual yoga poses, potentially informing the design of yoga-based rehabilitation protocols for different movement disorders and injuries.

% good thing to check is whether just upper extremity or lower or torso torques are correlated with each other. 
% check the realtive order of magnitude in coefficients used

\bibliography{myReferenceList_Nov_11_2024}  
\end{document}